PAPER

# Operations Smart Contract to Realize Decentralized System Operations Workflow for Consortium Blockchain*

Tatsuya SATO†, *Nonmember,* Taku SHIMOSAWA†, *Nonmember and* Yosuke HIMURA††, *Member*

**SUMMARY** Enterprises have paid attention to consortium blockchains like Hyperledger Fabric, which is one of the most promising platforms, for efficient decentralized transactions without depending on any particular organization. A consortium blockchain-based system will be typically built across multiple organizations. In such blockchain-based systems, system operations across multiple organizations in a decentralized manner are essential to maintain the value of introducing consortium blockchains. Decentralized system operations have recently been becoming realistic with the evolution of consortium blockchains. For instance, the release of Hyperledger Fabric v2.x, in which individual operational tasks for a blockchain network, such as command execution of configuration change of channels (Fabric's sub-networks) and upgrade of chaincodes (Fabric's smart contracts), can be partially executed in a decentralized manner. However, the operations workflows also include the preceding procedure of pre-sharing, coordinating, and pre-agreeing the operational information (e.g., configuration parameters) among organizations, after which operation executions can be conducted, and this preceding procedure relies on costly manual tasks. To realize efficient decentralized operations workflows for consortium blockchain-based systems in general, we propose a decentralized inter-organizational operations method that we call Operations Smart Contract (OpsSC), which defines an operations workflow as a smart contract. Furthermore, we design and implement OpsSC for blockchain network operations with Hyperledger Fabric v2.x. This paper presents OpsSC for operating channels and chaincodes, which are essential for managing the blockchain networks, through clarifying detailed workflows of those operations. A cost evaluation based on an estimation model shows that the total operational cost for executing a typical operational scenario to add an organization to a blockchain network having ten organizations could be reduced by 54 percent compared with a conventional script-based method. The implementation of OpsSC has been open-sourced and registered as one of Hyperledger Labs projects, which hosts experimental projects approved by Hyperledger.
*keywords:* Consortium blockchain; Hyperledger Fabric; System operations; Blockchain network operations.

## 1. Introduction

Recently, enterprises have paid attention to consortium blockchains (BCs) like Hyperledger Fabric [5] for efficient inter-organizational business transactions. Different from public BCs like Bitcoin [3] and Ethereum [4], which consist of many and unspecified participants, consortium BCs allow only inter-authorized organizations (forming a consortium) to construct a limited transaction scope to achieve a high transaction performance. As applications of BCs have expanded from traditional cryptocurrencies to various forms of asset management in accordance with the recent capability of BCs in dealing with "smart contracts (SCs)." In the context of BCs, SC is user-defined logic for business contracts and transactions, which are automatically executed over a BC network based on a distributed consensus protocol. Consortium BCs with SC features are expected to enable decentralized business transactions among multiple companies or organizations efficiently in various industries.

Toward realizing production uses of consortium BC-based systems, it is necessary to establish system management and operations for BC-based systems. In general, system management is a range of work done to keep systems running stably; system operations are tasks done for system management. Examples of system operations include installing software, updating the version of installed software, adding a new server for a new user, booting or halting service processes, creating backups, and restoring from them.

We consider that establishing decentralized system management for consortium BC will be essential to maintain the value of introducing consortium BCs. We assume that a single consortium BC-based system can be built across organizations. In such a system, an administrator in one organization does not have access permissions for the nodes for another organization. If a single administrator operates all nodes, the administrator becomes the single point of trust in the BC network or has excessive access permissions. If each organization's administrator operates their own nodes, a gap of operations (e.g., different configurations) could happen and, it may prevent the system from working. In summary, the problem is that it is difficult to execute inter-organizational operations in a decentralized and efficient manner.

Decentralized system operations have recently been becoming realistic with the evolution of consortium BCs, e.g., BC network-level operations can be partially executed in a decentralized manner according to Hyperledger Fabric v2.x, which has been released since 2020. Hyperledger Fabric (or just "Fabric") is one of the open-source software (OSS) platforms for consortium BCs that has been attracting attention in the enterprise field. Fabric has become production-ready and is mature as the recent LTS (Long-Term Support) version, v2.2, has been released. In Fabric

---

†The authors are with Hitachi Ltd., Kokubunji-shi, 185-8601, Japan.
††The author is with Hitachi America Ltd, Santa Clara, CA, 95054, USA.

v2.x, the individual operational tasks for the BC network operations, such as command execution of configuration change of channels (Fabric's sub-networks) and upgrade of chaincodes (Fabric's smart contracts), have been refined and can be executed in a decentralized manner. The BC network operations are performed by the combination of these individual tasks with organizations cooperating with each other.

Regarding Hyperledger Fabric, there is still a remaining issue that operations workflow should be decentralized and systematized from the end-to-end view. Indeed, the operations workflow also includes the preceding procedure of sharing, coordinating, and pre-agreeing the operational information (e.g., configuration parameters) among organizations, after which operation executions can be conducted. We call an "end-to-end" operations flow the sequence of tasks from the preceding procedure to the operation executions. This preceding procedure is not supported by Fabric and currently relies on costly manual tasks outside the BC.

To solve the problem of efficient decentralized operations for consortium BC-based systems in general, we propose an inter-organizational operations execution method, which we call Operations Smart Contract (OpsSC). This method defines system operations as SCs and makes it possible that inter-organizational operations can be performed (1) without relying on decisions by a specific organization (2) with uniform procedures and configuration parameters (3) efficiently.

As a practical application of the proposed method, we design and implement OpsSC for BC network-level operations workflow with Hyperledger Fabric v2.x. Specifically, the OpsSC focuses on BC network operations regarding operating channels and chaincodes, which are essential for managing Fabric's BC networks, and we deal with Fabric v2.2, which is the latest LTS version for this present work. We clarify detailed workflows of those operations from the view of OpsSC framework so that those operations can effectively be implemented with OpsSC.

The main contributions of our research are as follows:
- The idea to realize decentralized operations for BC-based systems by using BC-native features such as SC and BC consensus protocol.
- Practical application of the OpsSC concept and decentralization of end-to-end BC network operations workflows for Hyperledger Fabric: For the BC network operations, especially chaincode and channel operations, which are essential of Hyperledger Fabric operations, we implemented an OpsSC to support the end-to-end operations workflows from pre-agreement, operational information sharing to operations execution. After the initial setup of the BC network, the OpsSC allows the typical BC network operations to be executed in a decentralized manner.
- Quantitative evaluation of the OpsSC: We modeled typical scenarios for BC network operations in Hyperledger Fabric and evaluated the effect of reducing the operational costs by using OpsSC based on the model.

In this paper, we first show the general problems of system operations for consortium BC-based systems in Sec. 2 and then present the concept of OpsSC, a proposed method to solve the problems in Sec. 3. In the following sections, we move on to Hyperledger Fabric-specific topics. Sec. 4 describes issues on end-to-end BC network operations workflows for Hyperledger Fabric-based systems, and Sec. 5 presents the design and implementation of OpsSC to apply the end-to-end operations workflows. Then, Sec. 6 shows the evaluations of OpsSC, especially the quantitative evaluations by scenario-based operational cost estimation. Sec.7 shows discussions on the research. Finally, We show related work in Sec. 8 and conclude this research in Sec. 9.

## 2. General Problem about System Operations for Consortium BC-based Systems

2.1 Consortium BC-based Systems

A system using consortium BC like Hyperledger Fabric (referred to as consortium BC-based system) is typically a system consisting of (a) BC platform parts (mainly nodes like peers and orderers) and (b) application parts (including SCs). Fig. 1 shows the assumed BC-based system configuration in this research. In the production phase, each organization participating in the BC network owns nodes. In such system configuration, each organization has separated administrators. There may be multiple administrators in an organization depending on the system scale and management scheme.

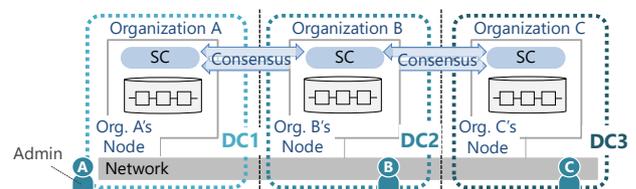

**Fig. 1** Architecture of consortium BC-based systems.

2.2 System Operations for Consortium BC-based Systems

In the production phase, system operations are necessary. In general, system operations are tasks executed by administrators to maintain the system.

For BC-based systems, there are two types of system operations:
- **Single-organizational operations:** Operations closed to one organization
  - e.g., modifying the frontend for each org's portal
- **Inter-organizational operations:** Operations that need to collaborate with other organizations specialized for BC-based systems
  - e.g., deploying the same SC in the same period over

the BC network

2.3 General Problem about System Operations for Consortium BC-based Systems

There are a lot of conventional management tools like job management servers, Infrastructure as Code (IaC) tools, and so on. They enable administrators to do general (means single-organizational) operations efficiently. However, they do not cover inter-organizational ones.

Fig. 2 shows probable ways to do inter-organizational operations. The first way is that a single administrator operates all nodes (Way 1). However, this way has a problem (Problem 1). One reason is that the administrator becomes the single point of trust (SPOT) in the BC network. Also, the administrator cannot access nodes owned by other organizations because of a lack of permissions. The second way is that each organization's administrator operates its own nodes (Way 2). In this way, a gap of operations could happen. Failing in consensus on the configuration can result in the failure of the whole system (Problem 2). The "gap of operations" concretely means that either operational procedures, timing, or configuration parameters are not consistent across organizations.

- **Operational procedure** is a step-by-step procedure for executing a system operation (e.g., a set of executed operational commands and scripts).
- **Timing** is a planned and/or on-demand time at which to start executing each operation (e.g., for periodical or emergency maintenance).
- **Configuration parameter** is a value described in the procedure and assigned for each operation execution (e.g., command arguments such as version number for software update commands).

Considering updating the SC as an example of Problem 2, if the configuration parameters including the version of the SC and the execution timing of the update are not consistent across organizations, the SC may become temporarily unusable (in some organizations using the wrong version) or may not be updated to the new version in the BC network for a long period of time.

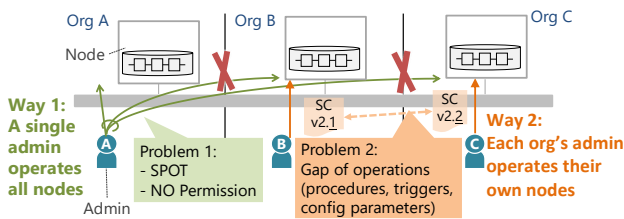

**Fig. 2**   Probable ways of inter-organizational operations and the problems.

## 3. Proposal: Operations Smart Contract (OpsSC)

### 3.1 Concept of OpsSC

To solve Problem 1 and Problem 2 described in Sec. 2.3, we propose a smart contract-based system operations method (we call the method "OpsSC"). Fig. 3 shows the conceptional diagram of the proposed method. The primary idea is to define system operations workflow as a SC. Operational procedures like commands and operations timing are defined as a SC (for system operations workflow referred to "OpWF as SC"). When the nodes receive an invocation transaction (TX) of the SC (Step 0 in Fig. 3), each node establishes consensus on each other (Step 1), and the nodes share configuration parameters and control workflow over the SC (Step 2). Then, each node executes operations based on the SC (Step 3). As a result, the operations are unified over multiple organizations. In summary, this method enables cross-organizational operations without SPOT or sharing credentials by using the consensus mechanism of BC. Also, it enables administrators to execute unified procedures with unified configuration parameters based on the SC.

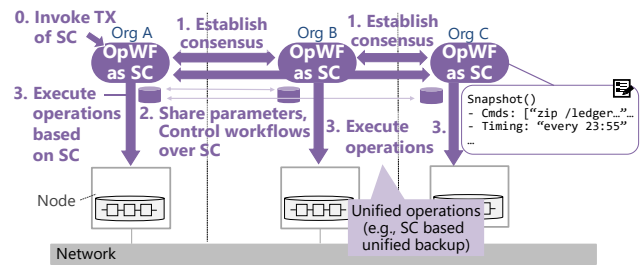

**Fig. 3**   Concept of OpsSC.

### 3.2 Extended Architecture with Operations Agents

Fig.4 shows an extended architecture of OpsSC with operations agents as one form of implementation. An operations agent is an independent daemon program that executes operations to nodes according to the status of the OpWF as SC. Typically each organization has its own agent to avoid SPOT, and that agent operates all nodes owned by that organization. Instead of executing operations to nodes directly, the OpWF as SC issues the operational instructions described in the SC as operations events to the operations agents (Step 3). The operations agent for each organization receives the event and then executes operations to corresponding nodes for the organization based on the event (Step 4). This architecture could avoid unpreferable direct I/O from inside of the SC to resources outside the SC, which may cause non-deterministic TX or layer violation considered as a bad manner in current general BC platforms.

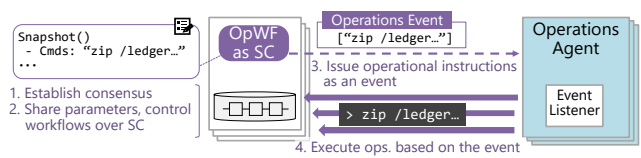

**Fig. 4** Extended architecture of OpsSC with operations agents

## 4. Hyperledger Fabric-based Systems and the Operational Issues

Sec. 3 presented the concept of OpsSC, a proposed method to solve the problem of inter-organizational system operations for consortium BC-based systems in general. From this section, we move on to Hyperledger Fabric-specific issues and solutions. Sec. 4 describes Hyperledger Fabric-based systems and the issues on end-to-end BC network operations workflows. Then, Sec. 5 presents the design and implementation of OpsSC to apply the end-to-end operations workflows.

4.1 Hyperledger Fabric-based Systems

Hyperledger Fabric is open-source platform software for consortium BCs that executes SCs (called chaincode or "CC"). Hyperledger Fabric stores all transaction histories in an append-only replicated ledger. Hyperledger Fabric consists of the following main components: peer, orderer, client, and Fabric CA (Certificate Authority).

Hyperledger Fabric introduces the execute-order-validate architecture. A client sends transaction proposals to the peers. Then, peers execute a transaction proposal and put a digital signature to the result. This step is also called "endorsement." After the execution of the transaction proposal, a client sends transactions to orderer nodes. Orderer nodes produce a totally ordered sequence of transactions grouped in blocks. These blocks are broadcast to all peers. Each peer validates the transactions with respect to the endorsement policy and the consistency of the states. After the validation, a block is committed to the ledger.

Hyperledger Fabric separates the trust model into two parts: transaction ordering and transaction validation. The trust model for ordering depends on the protocol which the BC system uses. For example, Raft protocol is used to tolerate crash faults. Also, protocols for tolerating byzantine faults will support in the near future (proposed and discussed in the community). Meanwhile, the trust model for transaction validation is flexible. Each system defines its own trust assumption by using the endorsement policy. Any transaction must satisfy the endorsement policy.

Fabric CA is a CA component to generate the certificates and keys to configure and manage identity in the BC network.

Each peer stores the ledger and the world state. The world state holds the current values of the result of reading/writing states based on the BC in a database named StateDB.

Hyperledger Fabric provides a feature to create sub-networks, called "Channels," under a BC network. Each channel may have different peers as its members (organizations). A channel has its own ledger separated from other channels. So, chaincodes are managed for each channel.

For more information about Hyperledger Fabric, refer to [5][6][7].

Fig. 5 shows assumed Hyperledger Fabric-based system configuration corresponding to Fig. 1. Each organization has and manages its own nodes. A BC network works by the interaction of the nodes that are owned by multiple organizations.

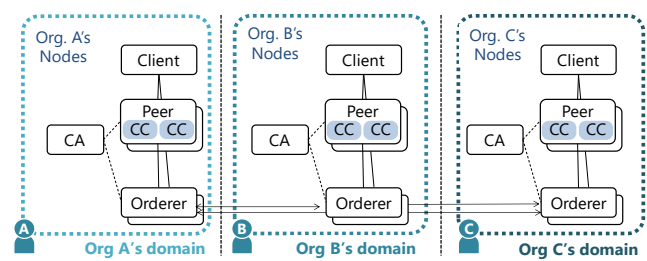

**Fig. 5** Architecture of Hyperledger Fabric-based systems.

4.2 BC Network Operations for Fabric and the Issues

**1) Overview**

BC network operations are common and essential operations of Hyperledger Fabric-based systems, and there are mainly two types: channel operations (e.g., adding operations to a channel) and chaincode operations (e.g., deploying a chaincode on a channel).

To execute their operations, Fabric provides commands (e.g., *peer* command) to control peer and orderer. In Fabric v2.x, which has been released since 2020, the commands have been refined, and the centralized parts are eliminated. To ensure that configuration values that need to be managed across organizations are not dependent on a specific organization, Fabric internally uses special chaincodes called "System Chaincodes (SCCs)," which makes it possible to share configuration values on-chain.

The typical BC network operations workflow in Fabric is shown in Fig. 6. As shown in the right part, the BC network operations are performed by the combination of the individual commands from each organization. As a mechanism to make it decentralized, these commands need to be executed by multiple organizations, and they need to be done in the proper order, and the configuration parameters (e.g., chaincode definition and source code information for chaincode deployment) must be coordinated across multiple organizations. Since such coordination is not supported by Fabric, it is necessary to share and adjust operational/configuration parameters among organizations in advance, outside of Fabric (i.e., off-chain), as shown in the figure on the left. For example, in chaincode deployment, each organization must approve the chaincode definition with the same parameters as the other organizations. So, organizations need to share and coordinate the configuration parameters which include the source code information and definition on the chaincode offline with other organizations in typical cases.

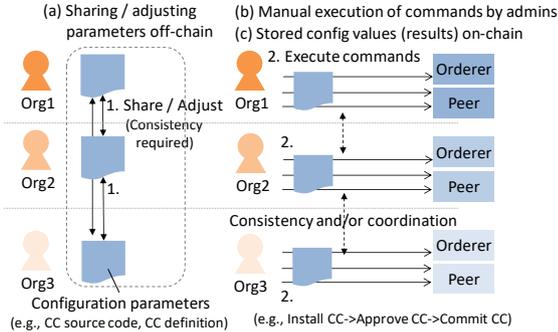

**Fig. 6**  Typical BC network operations workflow in Fabric v2.x.

In summary, such operations workflow also includes the preceding procedure of sharing, coordinating, and pre-agreeing the operational information (e.g., configuration parameters) among organizations, after which operation executions can be conducted. This preceding procedure is not supported by Fabric and currently relies on costly manual tasks outside the BC (i.e., off-chain).

In the following sub-sections, we analyze and describe the typical operation workflow and issues for each of channel and chaincode operations with the existing features of Fabric.

**2) Channel Operations**

Channel operations are for creating a new channel and updating a channel to add organizations and/or orderer nodes to the channel or to remove them from the channel. Channel operations are mainly executed by using multiple sub-commands of peer channel command. Channel configuration is stored in a collection of configuration transaction (configtx) per channel. Internally, Fabric utilizes the SCC called Configuration System Chaincode (CSCC). Channel configuration is updated/created via a configuration transaction. The typical operations workflow is shown in Fig. 7.

The series of steps is as follows:
1. Fetch the config block containing the channel configuration from the channel via peer and/or orderer
2. Modify the channel configuration and create ConfigUpdate (also called delta) that is the difference between the original and the updated configuration
3. Collect sufficient number of signatures (ConfigSignature) to the ConfigUpdate from each organization
4. Create an enveloped configuration transaction (configtx) based on artifacts (ConfigUpdate and ConfigSignatures) in Step 2 and 3 and send the configtx to the target channel via peer and/or orderer

As a remaining issue, the administrators need to share the artifacts with the other organizations out of Fabric.

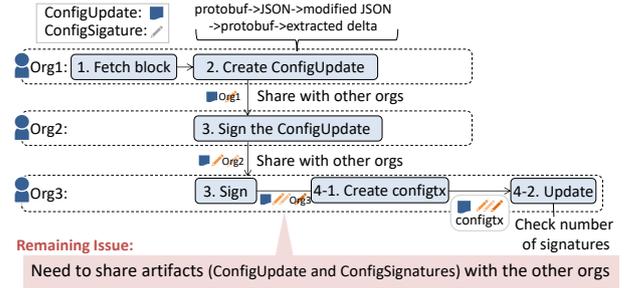

**Fig. 7**  Typical channel operations to update a channel.

We clarify a channel operation to add an organization to channels as the most typical case. The typical workflow of adding an organization is shown in Fig. 8.
The points of this typical workflow are as follows:
- In this channel operation, the inter-organizational operation is only updating channel configuration tasks (tasks with thick borders in the figure). The other tasks are single-organizational operations (in this example, performed by a newly added organization). The tasks with thick borders are the same as the flow shown in Fig. 7.
- Normally, when adding an organization, it is necessary to update the system channel (which is a special channel to manage the member information of "consortium organizations" which have orderer nodes) and each application channel.

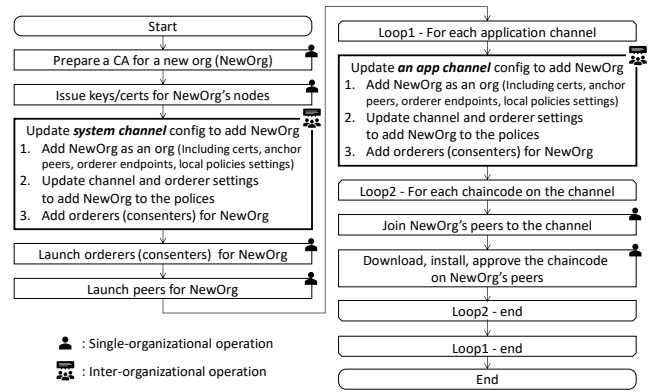

**Fig. 8**  Typical workflow to add a new organization to a BC network.

**3) Chaincode Operations**

Chaincode operations are mainly for deploying a new chaincode to a channel and upgrading a chaincode. For Fabric v2.x, chaincode operations are mainly executed by using multiple sub-commands of peer lifecycle chaincode command. Internally, it is implemented as a SCC (_lifecycle in v2.x series). In the v1.x series, there was still a centralized process, so some configuration parameters were determined by specific organizations (i.e., SPOT). In the new chaincode lifecycle from v2.0, a new step for approving chaincode definition from each organization was added. It can eliminate centralized processes in deploying and upgrading chaincode.

The typical operations workflow is shown in Fig. 9.

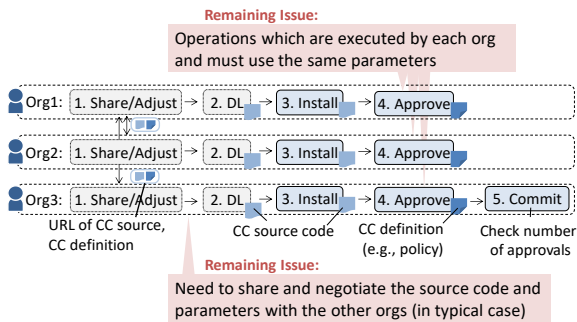

Fig. 9 Typical chaincode operations in Hyperledger Fabric v2.x.

The series of steps is as follows:
1. Organizations share/adjust the chaincode source code and the chaincode definition (e.g., endorsement policy) of the target chaincode outside of Fabric. Repositories such as Git may be used to share the source code.
2. Each organization downloads the chaincode source code outside of Fabric.
3. Each organization packages the chaincode. Then, each one installs the packaged chaincode on its peers.
4. Each organization approves the chaincode definition with respect to its own organization.
5. One of the organizations commits the chaincode definition to a channel. For that command to succeed, the number of organizations with an exact match of the approved definition must satisfy the chaincode update policy.

To summarize the above workflow issues, first, the workflow includes operations that are executed by each organization and must use the same parameters. Also, administrators need to share and negotiate the source code and configuration parameters with the other organizations in the typical case.

4.3 Target Scope of OpsSC for Hyperledger Fabric

Since these costly tasks described in Sec. 4.2 are always incurred for chaincode/channel operations, the issue must be common and important for Fabric-based systems. Therefore, a mechanism like OpsSC should be applied to these operations.

To improve the issue, we design and implement an OpsSC for Hyperledger Fabric v2.x, which is essential for the BC network operations: chaincode operations and channel operations. We target Fabric version v2.2, which is the latest LTS version at the time of this research. As shown in Fig. 10, the OpsSC helps administrators negotiate config parameters for operations and execute the operations to nodes automatically and in a decentralized manner.

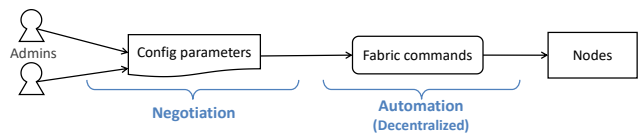

Fig. 10 Target area of OpsSC for Hyperledger Fabric.

In order to make the difference between pure Fabric and the OpsSC realized in this research clearer, the comparison is shown in Table 1. The OpsSC expands the scope of what can be managed on-chain compared with pure Fabric. Managing information on-chain means, in other words, systematizing workflows related to that information so that they are not dependent on a specific organization.

Table 1 Comparison between Pure Fabric and OpsSC.

| Category | Item | Pure Fabric | OpsSC for Fabric |
|---|---|---|---|
| Information management | Sharing/adjusting operational/ configuration parameters | Off-chain | On-chain |
| | Storing configuration values | On-chain | On-chain |
| | Managing operational execution history | Off-chain | On-chain |
| Operation | Executing operational commands among multiple organizations | Manual | Automated |

## 5. OpsSC for Hyperledger Fabric v2.x

This section presents the design and implementation of OpsSC for Hyperledger Fabric v2.x, specifically for operating chaincodes and channels. This implementation has been open-sourced and registered as one of Hyperledger Labs [23] projects (which hosts experimental projects officially approved by Hyperledger)[1]. We present the basic design of the OpsSC (Sec. 5.1 (1)-(3)) and the detailed design and implementation (Sec. 5.2 (1)-(3)) to fit Fabric network operations based on the analysis described in Sec. 4.2 (2)-(3).

5.1 Basic Design

**1) High-level Design**

Fig. 11 shows a basic framework that we designed based on the OpsSC architecture that introduced operations agents for each organization. The OpsSC shares operational information and manages workflows on the chaincodes for OpsSC. In addition, (a) to share, coordinate, and pre-agree operational information among organizations on-chain, the OpsSC adds a proposal/voting function: an organization proposes an operation, and each organization votes on it. Furthermore, (b) agents for each organization automatically execute multiple commands to accomplish the operations based on the proposal and the instructions from the chaincodes. As shown in Sec.4.2, the workflows cannot be expressed in a simple sequence because they contain tasks that may differ from organization to organization (e.g., committing the chaincode definition by a single organization). So, (c) the chaincodes have the ability to control the complex flows based on the past execution results stored on-chain.

---
[1] https://github.com/hyperledger-labs/fabric-opssc

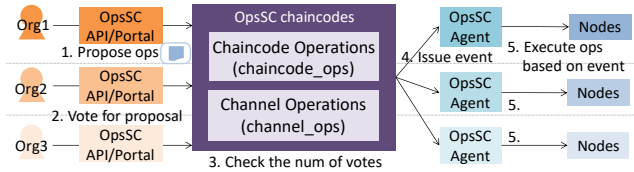

**Fig. 11** Basic Framework of OpsSC for Hyperledger Fabric.

This framework consists of three components: OpsSC chaincode, OpsSC API server, and OpsSC agent:
- *Chaincode* provides functions to manage operations workflows and issues operations events. An operations workflow is a workflow that represents what each administrator or organization must do to accomplish a certain operation. An operations event is information that represents the progress of the workflow and includes operational instructions on what each agent should do next. The operations events are implemented by Fabric's Chaincode Event feature.
- *API server* provides REST API for each organization's administrator to interact with the OpsSC chaincodes. It is implemented as a client program of Fabric. Typically, administrators access the API through their respective organization's portal.
- *Agent* for each organization executes operations based on the operations events to all nodes for the organization. It is also implemented as a client program of Fabric.

A typical flow for the framework is as follows:
1. An administrator for an organization requests a proposal of an operation (e.g., deploying a chaincode or updating a channel) via the organization's portal. The proposal is automatically shared on the OpsSC chaincode.
2. Administrators for other organizations approve (vote for) the proposal via each organization's portal.
3. The OpsSC chaincode checks the number of votes.
4. If a majority of the approvals is collected, the chaincode issues operations events for the proposal.
5. An agent for each organization executes operations to all nodes for the organization based on the operations events. The agent sends the operation results to the OpsSC chaincode. The results are stored on the OpsSC chaincode.

In the current implementation, OpsSC chaincodes are prepared for both channel and chaincode operations, and API servers and agents are equipped with functions for each OpsSC chaincode. Since these chaincodes also need to belong to a channel, they are deployed on a dedicated channel (called OpsChannel) prepared for OpsSC.

**2) Failure Models**

Agent failure (including malicious agent and operation failure by the agent) can be considered equivalent to peer failures. For operational commands that affect a single organization (e.g., approving the chaincode definition), the impact of the failure is closed within that one organization. For operational commands that affect the whole system (e.g., committing the chaincode definition, sending the configtx to the channel), the Fabric layer checks their validity according to the endorsement policy. As long as do not exceed what the endorsement policy is tolerant, the failure will not affect the system. Therefore, OpsSC can simply randomly select an organization to execute the commands, and if it fails, simply reselect the next organization.

**3) Common Functionalities**

Fig. 12 represents the functions (interfaces) that an OpsSC chaincode should have and the information that they should hold as (world) state in common:
- *Proposal* is a state that represents and manages a proposal of an operation. *Proposal* should contain *Configuration Parameters* and will contain *Artifacts*, which are intermediate deliverables, for the operation. In association with each *Proposal*, histories of task executions by administrators and agents of each organization (*Task Histories*) should be managed and stored as the states. The OpsSC chaincode manages *Status* of each *Proposal*, which is updated according to the internal state transition model, *Task Histories*, and controls the operational workflow according to *Status*.
- *Request()* is a function to request a proposal of an operation.
- *Vote()* is a function for each organization to vote on a proposal.
- *RegisterResults()* is a function to register the results of a task execution related to a proposal by an administrator or agent of each organization.

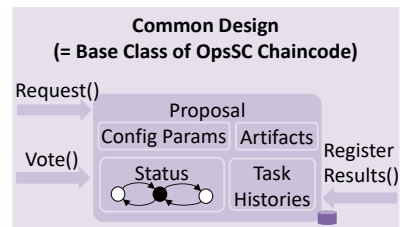

**Fig. 12** Common functionalities of OpsSC chaincodes.

Each OpsSC chaincode should have the above states and functions in common, although the names and implementation methods may change slightly.

5.2 Detail design and implementation

**1) Channel Operations**

OpsSC for operating channels streamlines channel updates across multiple organizations, such as shown in Fig.

7. This means that the OpsSC supports the workflow in the bolded part of Fig. 8, and excludes other parts.

Fig. 13 shows the overview of the OpsSC for operating channels. Note that this figure focuses on the parts of the framework described in Fig. 11 and 12 that are specific to this feature, and that the names may be a little different from the actual implementation to simplify the explanation. The OpsSC chaincode provides functionalities to share channel updates and signatures between different channel members. It provides SC functions to request a channel update proposal (that supports both creating and updating a channel), vote for the proposal by each organization with the signature, and register the status of operations to the proposal by each agent. The proposal format is designed as a human-readable format. Also, it provides SC functions to put/get information on channels (including the joining organizations) because Hyperledger Fabric does not currently provide a way to obtain the list of existing channels.

Because the original Fabric does not provide an official utility tool to create ConfigUpdate (Step 2 in Fig. 7), the process for channel configuration updates is both tedious and error-prone. We implement and use a generic configtx generation tool using Config Transaction Library, which is a standalone library provided by Fabric to create and modify a configtx. This tool enables to perform basic operations to a channel and output the results as a configtx. It is assumed that this tool is not only used directly by the administrators but also used internally by other tools such as OpsSC.

A typical flow for the OpsSC is as follows:

1. An organization creates a human-readable channel update proposal and requests it to the OpsSC. Internally, the OpsSC converts the proposal to ConfigUpdate and ConfigSignature with Config Transaction Library.
2. Other organizations vote for the proposal shared on the OpsSC. Internally, the OpsSC creates ConfigSignature for the proposal from the voting organization with Config Transaction Library.
3. When the majority of votes are collected, one of the agents automatically updates the channel with the configtx based on the proposal and votes.

**2) Chaincode Operations**

The OpsSC for operating chaincode streamlines the end-to-end chaincode deployment/upgrade operations, such as shown in Fig. 9.

Fig. 14 shows the overview of the OpsSC for operating chaincodes. Note that this figure focuses on the parts of the framework described in Fig. 11 and 12 that are specific to this feature, and that the names may be a little different from the actual implementation to simplify the explanation. The chaincode provides functionalities to communicate information about chaincode source code and chaincode definitions to be deployed between different channel members. It also provides SC functions to request a chaincode update proposal (that supports both deploying a new chaincode and upgrading a chaincode), vote for/against the proposal by each organization, and register the status of operations to the proposal by each agent. The chaincode internally calls the SC functions in the OpsSC chaincode for operating channels to get the information of the members of the channel that the proposed chaincode is deployed. The current implementation manages the chaincode source code as a link (including the URL and the commit ID) to the source code repository on the proposal instead of the source code itself and downloads the chaincode from there. This method has the advantage of saving the data capacity of the states, but if stricter decentralized management is needed, the source code can be managed directly as states.

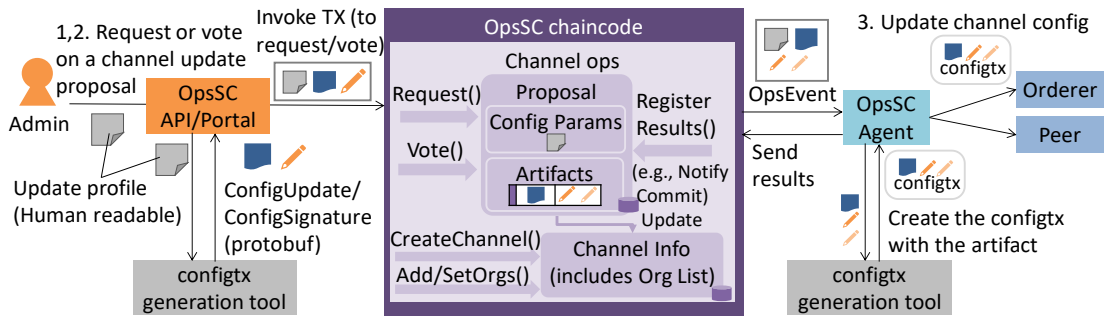

**Fig. 13** OpsSC for operating channels.

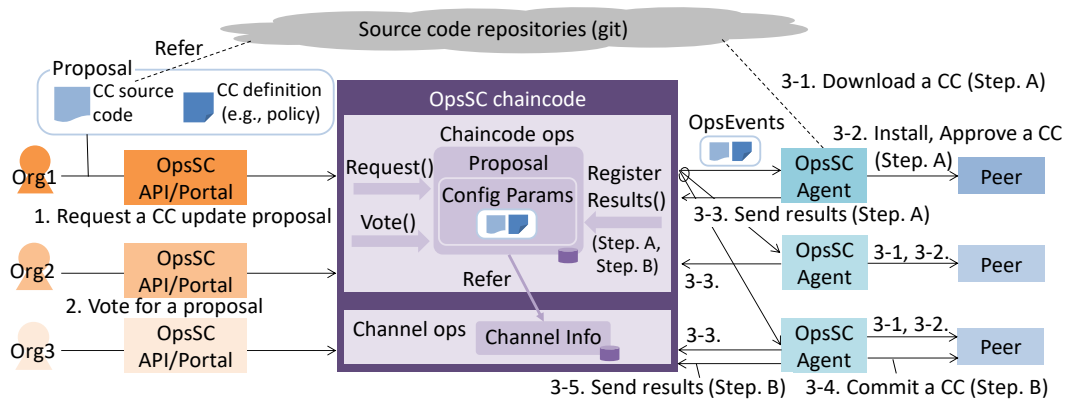

**Fig. 14** OpsSC for operating chaincodes.

A typical flow for the OpsSC is as follows:

1. An organization creates a proposal with a chaincode source code and a chaincode definition.
2. The other organizations vote for the proposal.
3. When the majority of votes are collected, the agents automatically deploy the chaincode based on the proposal, with downloading, installing, approving, and committing the chaincode.

In Step 3, the agent for each organization sends the operation results to the OpsSC chaincode, and the OpsSC chaincode manages the state of the operations workflow based on the operation results automatically. This makes it possible to automate the process of approval by each organization and commit by one organization.

**3) Organization Bootstrapping**

The OpsSC for operating channels only supports the operations workflow in the tasks with thick borders in Fig. 8. Therefore, after adding a new organization using OpsSC, the new organization will have to perform the organization bootstrapping steps manually. The steps are very complicated and the main information needed will be already managed in the OpsSC chaincodes for operating chaincodes and channels.

Therefore, we add a feature that allows the agent to automate even the above steps. When the agent for a new organization is launched or receives a chaincode event that notifies channel configuration updates, the agent automatically executes the following with referring to the OpsSC chaincodes to get the necessary information:

・ Joins their peers to OpsChannel
・ Deploys initial OpsSC chaincodes embedded as local files on their peers
・ Joins their peers to all the application channels which are managed by OpsSC
・ Deploys all the existing chaincodes (only the latest version of each) which are managed by OpsSC on their peers

With this feature, a new organization only needs to initially launch its own nodes (peers/orderers) and an agent, and the capability of the agent will allow it to automatically catch up with the latest status of the BC network.

## 6. Evaluations

6.1 Operational Cost Estimations

In order to evaluate the effectiveness of the OpsSC for Hyperledger Fabric implemented in this research, we evaluate the effect of reducing the operational cost of typical BC network operations with the proposed method (using OpsSC) compared with a conventional method (basic script-based operations). This evaluation was based on estimation with a cost model in typical BC network operation scenarios.

**1) Cost Model**

We model the typical procedures of the conventional and proposed methods in the following two operational scenarios for existing BC networks:

- **Scenario 1 (Channel operations):** Add a new organization (with peers and orderers) to all channels
- **Scenario 2 (Chaincode operations):** Deploy new chaincodes to a channel

The conventional and proposed methods are as follows:

- **Conventional method (Script-based Operations).** In the method, the administrators execute the operational steps manually. Note that it assumes that each step is organized into reasonable units and automated by scripts. Therefore, it is as efficient as possible compared to the actual simple manual operations.
- **Proposed method (using OpsSC).** In the method, the administrators execute the operation steps using the OpsSC for Fabric, as shown in the previous section. For the steps that are not covered by the OpsSC, we assume that the same scripts as in the conventional one are used.

In order to conduct a quantitative evaluation, we define a cost model. "operational costs" here means the man-hours or elapsed time required for operational work. The operational cost is evaluated from the following two perspectives:

- **Total Operational Cost (TOC):** as a perspective on man-hours. It is the total operational cost for all the steps executed by the administrators of all organizations. This is the operational cost required for executing a scenario in the BC network.
- **Lead Time (LT):** as a perspective on elapsed time. It is the end-to-end time it takes to complete the scenario, i.e., the time from starting executing the first step to finishing executing the last step.

We focus on the comparison of cost estimates and assume that the execution time of each step would be almost uniform due to scripting. For each step, the unit cost is set to "1" for each number of executions. We believe that even normalized costs can be a meaningful performance metric, even though it is a simplified model, because we assume the following situations in this paper:

- **Regarding TOC:** We consider that, by automation

through scripting (conventional method) or SC (proposed method), the cost required for each step (as manual tasks) can be summarized mainly into the tasks of inputting the config parameters for the step (transcribing information from the design documents or proposals by other organizations) and the tasks of communicating the execution results of the step to the next step. Since the cost required for these summarized tasks would not be expected to differ significantly between steps, the unit cost of each step is assumed to be constant.

- **Regarding LT:** We consider that each step can be summarized as the processing time required for automatic execution by scripts or SC, in addition to the time required for manual tasks for the summarized tasks in the TOC. In the cases of chaincode/channel operations, the time required for manual tasks (minutes or more) is more dominant than the time required for automatic execution processing (about seconds or tens of seconds). How many manual steps are required can represent the cost effectively. So, the unit cost of each step is assumed to be constant.
- **Regarding the granularly of step division:** The steps are divided into units for different executors or for different configuration parameters that are given manually. In addition, the steps are divided if they are required to wait for results from other organizations. In the case of OpsSC, the results of other organizations are automatically shared over the organizations, so such steps are integrated as one.

**[Scenario 1: Adding a new organization]**

The operational steps and their costs are shown in Table 2. The parameters used in this cost calculation formula are as shown in Table 3. In the conventional method, Steps 1-A-1 to 3 are the steps for preparing the certificates and keys for the new organization's nodes, and the new organization executes them. Steps 1-A-4 to 5 are the steps to create configtx to add the new organization and share it to other organizations, and one existing organization will execute them. These steps need to be executed for all the channels, i.e., the application channels and the system channel in the BC network. Therefore, in TOC, those costs are $CH+1$. Steps 1-A-6 to 7 are the steps to collect signatures from the remaining existing organizations. In the scenario, the channel update policy is set to the majority of the participating organizations, which is the default setting. Therefore, in TOC, the costs are multiplied by $N/2$ as the minimum value. Step 1-A-8 is the execution of the channel update command, which should be executed by an existing organization for all channels. Steps 1-A-9 to 10 is the node launch of the new organization, where an existing organization passes the genesis block to the new organization, and the new organization launches its own nodes. Steps 1-A-11 to 14 are for the new organization to join its own nodes to all channels and deploy the existing chaincode to all application channels in order to catch up with the latest status of the BC network. Since the existing chaincodes have already been approved and committed, only this new organization needs to go through the approval process for each chaincode. Therefore, these costs are multiplied by the $CC$.

In the proposed method, instead of Steps 1-A-4 to 8, an existing organization requests a proposal to add the new organization via the OpsSC portal, and the other organizations approve the proposal (Steps 1-B-4 to 5). It is assumed that the approval from the majority is required as in the conventional method. In the proposed method, the number of target channels is added by 1 in these costs because the BC network has OpsChannel to manage the OpsSC chaincodes. The catching up process in Steps 1-A-11 to 14 of the conventional method is not necessary because the OpsSC agent for the new organization does the process automatically. Instead, it is necessary to launch the OpsSC agent and API server for the new organization (Step 1-B-8).

Table 2   Operational steps and the costs for Scenario 1.

| Conventional method | | | | Proposed method | | | |
|---|---|---|---|---|---|---|---|
| ID | Step | Cost TOC | LT | ID | Step | Cost TOC | LT |
| 1-A-1 | A new org launches a CAs for the org | 1 | 1 | 1-B-1 | Same as 1-A-1 | 1 | 1 |
| 1-A-2 | The new org issues certificates/keys for their nodes with the CAs | 1 | 1 | 1-B-2 | Same as 1-A-2 | 1 | 1 |
| 1-A-3 | The new org sends MSP info (certificates etc.) to an existing org | 1 | 1 | 1-B-3 | Same as 1-A-3 | 1 | 1 |
| 1-A-4 | The existing org creates each configtx to add the new org to each channel | $(CH+1)$ | $(CH+1)$ | 1-B-4 | An existing org proposes to add the new org to each channel via the org's OpsSC portal | $(CH+2)$ | $(CH+2)$ |
| 1-A-5 | The existing org share the each configtx to each channel | $(CH+1)$ | $(CH+1)$ | - | - | 0 | 0 |
| 1-A-6 | Each of the remaining existing orgs signs on to the each configtx | $(CH+1)*(N/2)$ | $(CH+1)$ | 1-B-5 | Each of the remaining existing orgs votes for (approves) the proposal via each org's OpsSC portal | $(CH+2)*(N/2)$ | $(CH+2)$ |
| 1-A-7 | Each of the remaining existing orgs shares each configtx signed by the org to the other orgs | $(CH+1)*(N/2)$ | $(CH+1)$ | - | - | 0 | 0 |
| 1-A-8 | An existing org updates each channel by using the configtx | $(CH+1)$ | $(CH+1)$ | - | - | 0 | 0 |
| 1-A-9 | An existing org shares the system genesis block to the new org | 1 | 1 | 1-B-6 | Same as 1-A-9 | 1 | 1 |
| 1-A-10 | The new org launches their nodes (peers and orderers) by using the genesis block | 1 | 1 | 1-B-7 | Same as 1-A-10 | 1 | 1 |
| - | - | 0 | 0 | 1-B-8 | The new org launches their OpsSC agent and API server | 1 | 1 |
| 1-A-11 | The new org joins the peers to each channel | $(CH+1)$ | $(CH+1)$ | - | - | 0 | 0 |
| 1-A-12 | The new org downloads the all existing chaincodes for each application channel | $CH*CC$ | $CH*CC$ | - | - | 0 | 0 |
| 1-A-13 | The new org packages/installs the all chaincodes to the peers for each application channel | $CH*CC$ | $CH*CC$ | - | - | 0 | 0 |
| 1-A-14 | The new org approves the chaincode definitions of the all chaincodes for each application channel | $CH*CC$ | $CH*CC$ | - | - | 0 | 0 |
| Total | | $CH*(3CC+N+4)+N+9$ | $CH*(3CC+6)+11$ | Total | | $CH*(N/2+1)+N+8$ | $2CH+10$ |

The costs of TOC and LT differ in the hatched areas in the table. For steps executed by multiple organizations, the TOC cost is accumulated by the number of organizations, while the LT cost is estimated for one organization because it can be executed in parallel.

Table 3  Parameters of cost models.

| Parameter | Description |
|---|---|
| $N$ | Number of existing organizations in the BC network |
| $CH$ | Number of existing application channels |
| $CC$ | Number of existing application chaincodes |

[Scenario 2: Deploying application chaincodes]

The operational steps and their costs are shown in Table 4. The parameters used in this cost calculation formula are as shown in Table 3. To deploy a chaincode to a channel, in the conventional method, one existing organization shares the source code and definition of the chaincode with other organizations (Step 2-A-1). Then, every existing organization needs to download that chaincode source code to a local space, install it in their own peers, and approve the chaincode definition (Steps 2-A-2 to 2-A-5). In the scenario, the chaincode deployment/upgrade policy is set to the majority of the approvals from participating organizations, which is the default setting. However, since the chaincode will be unavailable to organizations that have not approved it, it should be approved by all organizations in real usecases to maintain the availability of the BC network. Therefore, the cost of TOC is $N$. Step 2-A-5 is committing the chaincode definition that has been approved, and can be done by one organization.

Table 4  Operational steps and the costs for Scenario 2.

| Conventional method | | | | Proposed method | | | |
|---|---|---|---|---|---|---|---|
| ID | Step | Cost TOC | LT | ID | Step | Cost TOC | LT |
| 2-A-0 | (An existing org develops a new CC) | [Out of scope] | | 2-B-0 | (Same as 2-A-0) | [Out of scope] | |
| 2-A-1 | The existing org share the source code and definition of the new CC to the other orgs | 1 | 1 | 2-B-1 | The existing org proposes to add the new CC via the org's OpsSC portal | 1 | 1 |
| 2-A-2 | Every existing org downloads the new CC | $N$ | 1 | 2-B-2 | Each of the remaining existing orgs votes for (approves) the proposal via each org's OpsSC portal | $N/2$ | 1 |
| 2-A-3 | Every existing org packages/installs the new CC to the peers | $N$ | 1 | - | | 0 | 0 |
| 2-A-4 | Every existing org approves the definition for the new CC | $N$ | 1 | - | | 0 | 0 |
| 2-A-5 | An existing org commits the definition for the new CC | 1 | 1 | - | | 0 | 0 |
| Total (for 1 CC) | | $(3N+2)$ | 5 | Total (for 1 CC) | | $(N/2)+1$ | 2 |
| Total (for multiple CCs) | | $CC*(3N+2)$ | $5CC$ | Total (for multiple CCs) | | $CC*\{(N/2)+1\}$ | $2CC$ |

In the proposed method, one organization requests a proposal to deploy a chaincode, and the other organizations approve the proposal (Steps 2-B-1 to 2). After the number of approvals satisfies the condition, the agents automatically deploy the target chaincode to peers in all organizations. Therefore, in contrast to the conventional method, the cost of Step 2-B-2 is $N/2$.

The difference between the cost TOC and LT is the same as in Scenario 1.

2) Estimations

 [Scenario 1: Adding a new organization]

The results for TOC are shown in Fig. 15 left, where the parameters $CC$ and $CH$ are fixed at 2 and $N$ is varied from 2 to 20. Also, the results for LT are shown in Fig. 15 right, where the parameter $CC$ is fixed at 2 and $CH$ is varied from 2 to 10.

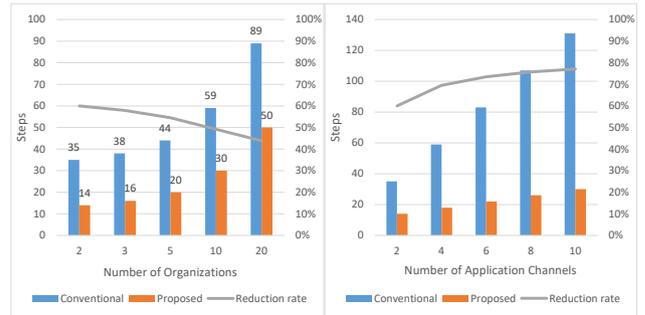

Fig. 15  Results (Left: TOC, Right: LT).

[Scenario 2: Deploying application chaincodes]

The results for TOC are shown in Fig. 16 left, where the parameters $CC$ and $CH$ are fixed at 2 and $N$ is varied from 2 to 20. Also, the results for LT are shown in Fig. 16 right, where the parameter $CC$ is varied from 2 to 10.

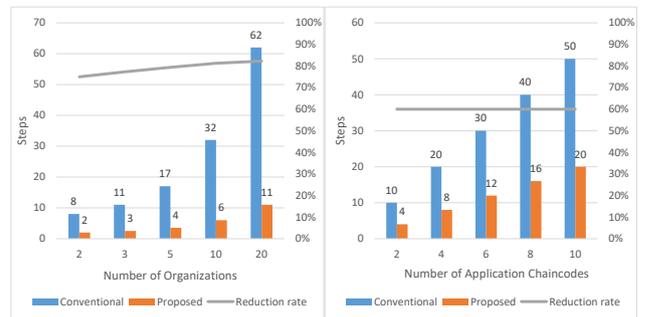

Fig. 16  Results (Left: TOC, Right: LT).

3) Discussions

Fig. 15 shows that both TOC and LT results have significant reduction effects in Scenario 1, and that the effect of TOC gradually decreases as the number of organizations increases. As the number of organizations becomes more dominant, the effectiveness converges to about 33 percent. According to the article [24], participating organizations to a consortium in the main practical use cases range from a few organizations to about 100 organizations. The above shows that the proposed method is effective in the number of organizations in practical use cases and can achieve a certain level of reduction no matter how much the number of organizations increases. Fig. 16 shows that both TOC and LT results have significant reduction effects in Scenario 2, especially for TOC. These results indicate the OpsSC is effective in reducing operational costs for both channel and chaincode operations.

In this evaluation, in order to reduce the dependence on specific projects as much as possible, we estimated the effect by accumulating the number of executions of each step. If the actual execution time of each step is known, it can be applied to improve the accuracy of the estimation. In real operations, each step needs to wait for the completion of the previous one. Therefore, especially for steps that are executed in parallel by multiple organizations, there is an overhead of waiting for completion by all of them. As the number of organizations increases, the variance in execution time of each organization increases. It may have a significant impact on LT. Considering the above, the difference in operational costs between the conventional and proposed methods can be even larger.

6.2 Qualitative evaluations

**Workflow automation.** End-to-end workflows are automated and ensured to be uniform by the OpsSC. Since the internal information transfer is also done automatically, the operational errors that could occur in the sharing of operational information and the transcription of parameters at the time of command execution by each organization in the conventional method can be prevented. By automating the execution timing, it is possible to reduce the variation of the start time and execution time of a task in each organization.

**Auditability.** The OpsSC records the approval history and operation execution history of the operation on-chain. These histories can be used as operational logs and evidence for auditing and other purposes.

**Capability of decentralized operations.** The operations workflow management by the OpsSC enables decentralized operations workflow for chaincode/channel. Once the initial setup is done, the typical chaincode and channel operations can be covered by OpsSC. In addition, the OpsSC can also perform the operation and management of the OpsSC itself (e.g., update of OpsSC) in a decentralized manner. Therefore, the OpsSC is a method that maintains the value of introducing consortium BCs.

## 7. Discussions

**Limitations of the current design and implementation.** The current implementation is specialized for channel and chaincode operations and cannot be used as is for other operations such as taking system-wide ledger snapshots, unified log settings for all organizations, checking peer versions across the organization and so on.

Although the current implementation itself is difficult to reuse, we believe that the basic design (framework and common functions) can be adapted to those other operations. The simple solution would be to implement an OpsSC for each new operation based on the basic design, but this may still be costly to develop. As we mentioned, channel and chaincode operations have very complex workflows, while the other possible operations so far do not seem to have that much complexity. Therefore, we consider that one idea is to design and implement a templated OpsSC that can be applied to various operations that do not have complex workflows.

## 8. Related Work

**Operation and management of Hyperledger Fabric-based systems.** Hyperledger Fabric introduces a special SC called "System Chaincode (SCC)," which makes it possible to run SCs in processes and is currently used for internal processing and configuration-value sharing on the BC platform (e.g., _lifecycle to manage chaincode lifecycle, CSCC to handle changes to a channel config). OpsSC for Fabric internally uses SCCs to operate the BC network. Fabric interop working group [22] aims to promote the interoperability of Fabric network service. It focuses on a scenario that a new organization joins a running Fabric network. The approach is to create artifacts for the join request with "Consortium Management Chaincode (CMCC)." The concept is very similar to ours, although the scope is different. OpsSC could be positioned as a form or application of the CMCC. Compared to the original CMCC, OpsSC has expanded its coverage to include end-to-end processes and is more automated. Since OpsSC could cover the operations of an entire BC-based system including the platform, it would complement the coverage of their two works and contribute to maintaining the quality of a system using Fabric.

**System operations and management as code.** Tools for practicing Infrastructure as Code (IaC), such as Ansible [8], Chef [9], and Terraform [10], have been spreading to enterprise fields, which can automatically and uniformly manage and provision an IT infrastructure even on heterogeneous OSes through the abstraction and code of a domain-specific language. Furthermore, in recent years, a standard specification and related research for template-based representation for the configuration of a system running on an arbitrary cloud environment using a domain specific model have also appeared [11]. Incorporating these prior arts into our proposed method will make it possible to apply our proposed method to BC-based systems built on heterogeneous OSes and cloud environments. The proposed method could help to extend the scope of IaC's automation to the adjustment of execution timing and dynamic parameters for each execution without cross-organizational access violations.

**Operational procedures management.** There are researches on improving the efficiency of operational procedures management. [14] shows a method that extracts reusable procedure parts from documents of operational procedures to improve the efficiency of managing the procedures. [12][13] shows a method that discovers a history of operations by automatically analyzing raw system logs. [13] shows a method that extracts operations workflows by analyzing text-based working histories for trouble tickets. By incorporating these methods into OpsSC, cross-organizational operations could be managed and executed more efficiently (e.g., support for designing/defining operations workflows).

**Application of consortium BC.** Various use cases and applications of consortium BC have been proposed and discussed, and the results are being published as articles including papers such as [20] and [21]. Considering concrete system operations by utilizing knowledge on the practical applications described in these articles would help in polishing our proposal to make it more feasible for production uses. In addition, we could also consider our proposal (especially OpsSC and the portal application) as opening up a novel form of BC application. Our study could also contribute to the evolution of BC applications.

**Research on development of SCs.** There are researches on development of SCs such as improvement of SC productivity and quality. For instance, [15] reveals the new research directions about BC application development such as testing, software tools and so on. [16] tries to apply traditional software design patterns to BC applications. For other instances, [17] is a study on security risk analysis of SC and [18] and [19] are studies on formal verification of SC. The results of these research fields could also be utilized for reduction of security risk and improvement of development productivity of our proposed "OpsSCs."

## 9. Conclusions

In this paper, we proposed an operations execution method for consortium BC-based systems named OpsSC (Operations Smart Contract). The primary idea is to define operations as SC so that decentralized inter-organizational operations can be executed effectively. Furthermore, we designed and implemented an OpsSC for BC network operations of Hyperledger Fabric v2.x, specifically for operating chaincodes and channels. The OpsSC helps administrators negotiate config parameters for operations and execute the operations to nodes automatically and in a decentralized manner. The current implementation has been open-sourced and registered as one of Hyperledger Labs projects. A cost evaluation using model-based estimation showed that the total cost of operations could be drastically reduced compared with a conventional script-based method. In this paper, the implementation of OpsSC focused on major blockchain operations, especially for chaincodes and channels, but there are other operations, such as taking a snapshot of ledger data. In the future, we will continue to enhance the OpsSC for Hyperledger Fabric to support the other operations as well.

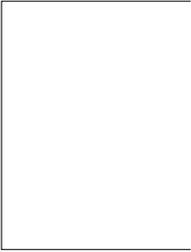

**Tatsuya Sato** received the M.S. degree in engineering at the University of Tsukuba, Tsukuba, Japan in 2009. He is presently with Center for Technology Innovation – Digital Platform, Hitachi, Ltd., Japan. His research interests include blockchain platforms and system operations. He is also a contributor for Hyperledger Fabric. He is a member of IPSJ.

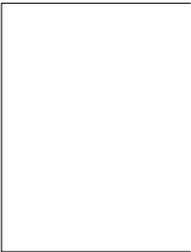

**Taku Shimosawa** received his Ph.D. in Information Science and Technology from the University of Tokyo, Japan in 2012. He is currently a chief researcher at Hitachi, Ltd. His research interests include operating systems, systems software and blockchain platforms.

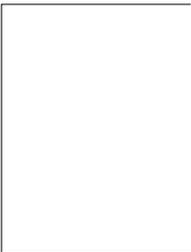

**Yosuke Himura** received his Ph.D in information and communication engineering at the University of Tokyo, Tokyo, Japan in 2016. He is presently with Hitachi America, Ltd. His research interests include computer network management and analysis.